\documentclass[12pt]{article}
\usepackage{amsfonts,amssymb,amsmath}
\usepackage[dvips]{epsfig}
\newcounter{tempeq}
\begin{document}
\title{Cat-States in the Framework of Wigner-Heisenberg Algebra}
\author{A. Dehghani$^{1}$\thanks{Email: a\_dehghani@tabrizu.ac.ir,  alireza.dehghani@gmail.com},
 \hspace{2mm}B. Mojaveri$^{2}$\thanks{Email: bmojaveri@azaruniv.ac.ir; bmojaveri@gmail.com},\hspace{2mm} S. Shirin$^{1}$\thanks{Email: siminshirin2000@gmail.com} and \hspace{2mm}M. Saedi$^{3}$\thanks{Email: m\_saedi@pnu.ac.ir}\\
{\small {\em $^{1}$Department of Physics, Payame Noor University,
P.O.Box 19395-3697 Tehran, I.R. of Iran\,}}\\
{\small {\em $^{2}$Department of Physics, Azarbaijan Shahid Madani
University, PO Box 51745-406, Tabriz, Iran \,}}\\
{\small {\em $^{3}$Department of Mathematics, Payame Noor University,
P.O.Box 19395-3697 Tehran, I.R. of Iran\,}}} \maketitle
\begin{abstract} A one-parameter generalized Wigner-Heisenberg algebra( WHA) is reviewed in detail. It is shown that WHA verifies the deformed commutation rule $[\hat{x}, \hat{p}_{\lambda}] = i(1 + 2\lambda \hat{R})$ and also highlights the dynamical symmetries of the pseudo-harmonic oscillator( PHO). \textbf{The present article is devoted to the study of new cat-states} built from $\lambda$-deformed Schr\"{o}dinger coherent states, which according to the Barut-Girardello scheme are defined as the eigenstates of the generalized annihilation operator. Particular attention is devoted to the limiting case where the Schr\"{o}dinger cat states are obtained. Nonclassical features and quantum statistical properties of these states are studied by evaluation of Mandel's parameter and quadrature squeezing with respect to the $\lambda-$deformed canonical pairs $( \hat{x}, \hat{p}_{\lambda})$. It is shown that these states minimize the uncertainty relations of each pair of the $su(1,1)$ components.
\end{abstract}
\section{Introduction}
Optical Schr\"{o}dinger cat states of light, consisting of a superposition of two coherent states, are of great interest in quantum optics as they have potential applications in various domains such as continuous variable quantum computation \cite{Neergaard, Ralph1}, quantum error-correcting codes \cite{Cochrane, Gottesman}, fundamental testings [5- 10] and precision measurement \cite{Ralph2, Joo}. The main challenge in almost all these applications is to generate states whose amplitude is large enough to perform good quality operations \cite{ Ralph1, Lund}. In quantum optics, cat states are defined as linear superposition of two coherent states with opposite phase:\footnote{Subscript `Sch' refers to the Schr\"{o}dinger and
specifies a particular type of quantum states that are called the
canonical coherent states, too.}
\renewcommand\theequation{\arabic{tempeq}\alph{equation}}
\setcounter{equation}{0} \addtocounter{tempeq}{1}
\begin{eqnarray}
&&\hspace{-1.5cm}|Cat\rangle_{+}:=\frac{|\alpha\rangle_{Sch}+|-\alpha\rangle_{Sch}}{\sqrt{2(1+e^{-2|\alpha|^2})}}=
\sqrt{\frac{1}{\cosh(|\alpha|^{2})}}\sum^{\infty}_{n=0}\frac{\alpha^{2n}}{\sqrt{(2n)!}}|2n\rangle,\\
&&\hspace{-1.5cm}|Cat\rangle_{-}:=\frac{|\alpha\rangle_{Sch}-|-\alpha\rangle_{Sch}}{\sqrt{2(1+e^{-2|\alpha|^2})}}
=\sqrt{\frac{1}{\sinh(|\alpha|^{2})}}\sum^{\infty}_{n=0}\frac{\alpha^{2n+1}}{\sqrt{(2n+1)!}}|2n+1\rangle.
\end{eqnarray}
These states are often referred to as even and odd coherent states, respectively, first introduced by Dodonov et.al in 1974 \cite{Dodonov}. They are called even and odd coherent state since such superpositions involve only even and odd Fock states. They are two orthogonal and normalized states which include two kinds of interesting nonclassical features: the even coherent state $|Cat\rangle_{+}$ has a squeezing effect but has no anti-bunching effect, while the odd coherent state $|Cat\rangle_{-}$ has an anti-bunching effect but has no squeezing effect [16- 20]. For an enough large displacement parameters $\alpha$, the states $|Cat\rangle_{\pm}$ can be interpreted as the quantum superposition of two macroscopically distinguishable states, which are called Schr\"{o}dinger-cat-like-states \cite{Haroche}. A well-studied example is the case of a two-level ion trapped in an external harmonic field. It turns out that the stationary states of the center of mass motion (in the harmonic trap) of a laser-driven ion are even or odd coherent states \cite{Matos, Leibfried}. Another method suggested for the preparation of cat states involves the coupling of an optical coherent field with a Kerr nonlinear medium \cite{Yurke}. From algebraic point of view, even and odd coherent states can be considered as the Burut-Girardello's $su(1, 1)$ coherent states [19, 25-28], i.e.
\renewcommand\theequation{\arabic{tempeq}\alph{equation}}
\setcounter{equation}{0} \addtocounter{tempeq}{1}
\begin{eqnarray}
&&\hspace{-1.5cm}{\hat{a}}^2|Cat\rangle_{+}=\alpha^2|Cat\rangle_{even},\\
&&\hspace{-1.5cm}{\hat{a}}^2|Cat\rangle_{-}=\alpha^2|Cat\rangle_{odd},
\end{eqnarray}
where $\hat{a}$ denotes the boson annihilation operator.

\textbf{In recent years a lot of attention was paid to the extension and deformation of the boson oscillator algebra}. Most of deformations of a boson algebra have been accomplished until now [29- 33]. Some of them are constructed by using the Jackson's q- calculus \cite{Jackson}, while others are not. One of the most interesting algebra which is not related to the q-calculus is Wigner algebra \cite{Wigner}. In 1950, Wigner proposed the interesting question,\\\\``{\em Do the equations of motion determine the quantum mechanical commutation relations?}" \\\\ \textbf{and he found that a deformation of the Heisenberg algebra (called Wigner algebra) could be introduced, leaving the equations of motion unchanged in the case of a free particle or an harmonic oscillator. According to Wigner's new quantization method, a supplementary term $\lambda \hat{R}$ can be introduced in the modified commutation relation between the position and the deformed momentum operator, where $\lambda$ is a constant called Wigner parameter and $\hat{R}$ is the parity operator}. The coordinate representation for the deformed momentum operator has been found \cite{Yang}, which realizes Wigner algebra. Also, there has been further interest in extending algebras, or even more basically in extending differential operators by reflection operators [37- 42].

Wigner algebra is an obvious modification of the boson algebra. Indeed, it reduces to an ordinary boson algebra when Wigner parameter $\lambda$ becomes zero. It is well known that Wigner algebra is linked to the reduced part of the two-particle Calogero model( or PHO) [43- 45], where Wigner parameter is related to the Calogero coupling constant. \textbf{This system is exactly solvable, the potential being the sum of the harmonic potential, $\frac{1}{2}x^2$, and the inversely quadratic potential, $\frac{1}{2}\frac{\lambda(\lambda-1)}{x^2}$. This model was first proposed by Post in 1956 when he studied the one-dimensional many identical particles problem in the case of the pair-force interaction between particles \cite{post1}}. Since 1961 such a quantum system has been studied by many authors [46- 58]. For example, Landau and Lifshitz studied its exact solutions in three dimensions \cite{landau}. Recently, Sage has studied the vibrations and rotations of the pseudo-gaussian oscillator in order to describe the diatomic molecule \cite{sage}, in which he briefly reinvestigated some properties of the PHO to study the pseudo-gaussian oscillator. Advantages of the pseudo-harmonic potential have been considered for improvements in the conventional presentation of molecular vibrations \cite{sage1}. Hurley found that this kind of PHO interaction between the particles can be exactly solved by separation of variables when he studied the three-body problem in one dimension \cite{hurly}. A few years later, Calogero studied the one-dimensional three- and N-body problems interacting pairwise via harmonic and inverse square (centrifugal) potential \cite{caf, caf1}. On the other hand, this potential was generalized by Camiz and Dodonov et al. to the non-stationary (varying frequency) PHO potential \cite{camiz, dod, dod1}. In addition, such a physical problem was also studied in arbitrary dimension $D$ [60- 64]. Also, Dong et .al have studied its dynamical group in two dimensions \cite{dong}. \textbf{The reduced part of the two-particle Calogero-Sutherland model has attracted considerable interests [43, 66, 67] in connection with its $su(1,1)$ dynamical symmetry \cite{Perelemov1, Perelemov2, demir}}. It is of great interest in quantum optics because it can characterize many kinds of quantum optical systems \cite{Barut, Perelomov, Zhang}. In particular, the bosonic realization of $su(1,1)$ describes the degenerate and non-degenerate parametric amplifiers \cite{WODKIEWICZ, Deh}. \textbf{For this reason, the study of coherent states
for PHO is of great importance. It has been recently achieved in the framework of $su(1,1)$ Lie algebra \cite{Agarwal1, dehghani1, demir, Tavassoly1}}. Because of this important progress in the generation of coherent states and their combinations, we are interested in extending the notion of even and odd coherent states to potentials other than the harmonic oscillator. In particular, we will consider even and odd coherent states associated with the PHO. This paper is organized as follows. We review WHA and its connection to the PHO in sections. 2 and 3, respectively. In section. 4, we construct generalized even and odd coherent states associated to the PHO and study some of the nonclassical properties such as sub-Poissonian statistics and quadrature as well as higher order squeezing effects. Finally, in section 5, we present a summary and conclusion.
\section{A Short Review of Generalized Hermite Polynomials $H^{\lambda}_{n}(x)$}
Generalized Hermite polynomials, $H^{\lambda}_{n}(x)$, were introduced by Szego \cite{Szego} and discussed latter in details in Refs. \cite{Markett, Rosenblum, Alvarez} as
\renewcommand\theequation{\arabic{tempeq}\alph{equation}}
\setcounter{equation}{0} \addtocounter{tempeq}{1}
\begin{eqnarray}
&&\hspace{-29mm}H^{\lambda}_{2n}(x)=(-1)^{n}2^{2n}n!L^{\lambda-\frac{1}{2}}_{n}(x^2),\\
&&\hspace{-29mm}H^{\lambda}_{2n+1}(x)=(-1)^{n}2^{2n+1}n!xL^{\lambda+\frac{1}{2}}_{n}(x^2),\end{eqnarray}
where $L^{\mu}_{n}(x)$ denote the Laguerre polynomials with $\mu>-1$ \cite{abramo}. \textbf{They reduce to the ordinary Hermite polynomials when $\lambda$ vanishes, i.e. $H^{\lambda=0}_{n}(x) = H_{n}(x)$}. The generalized Hermite polynomials (3) are orthogonal with respect to the weight function $|x|^{2\lambda}e^{-x^2}, x\in \mathbb{R}$, i.e.
\renewcommand\theequation{\arabic{tempeq}\alph{equation}}
\setcounter{equation}{-1} \addtocounter{tempeq}{1}
\begin{eqnarray}
&&\hspace{-14mm}\int^{\infty}_{-\infty}{|x|^{2\lambda}e^{-x^2}H^{\lambda}_{n}(x)H^{\lambda}_{m}(x)dx}=2^{2n}\left[\frac{n}{2}\right]!\Gamma\left(\left[\frac{n+1}{2}\right]+\lambda+\frac{1}{2}\right)\delta_{n,m}.\end{eqnarray}
Here $[a]$ denotes the greatest integer not exceeding $a$. \textbf{The interest of the generalized Hermite polynomials
is twofold:  (i) correctly weighted, they form an orthonormal set of $L^{2}(\mathbb{R}, dx)$, (ii) one can build the Bose-like oscillator calculus in terms of these polynomials \cite{Rosenblum}. Indeed defining $\psi^{\lambda}_{n}(x)\equiv \langle x|n, \lambda\rangle$ as}
\renewcommand\theequation{\arabic{tempeq}\alph{equation}}
\setcounter{equation}{0} \addtocounter{tempeq}{1}
\begin{eqnarray}
&&\hspace{-24mm}\psi^{\lambda}_{2n}(x):=\frac{|x|^\lambda
e^{-\frac{x^2}{2}}H^{\lambda}_{2n}(x)}{\sqrt{2^{4n}
n!\Gamma(n+\lambda+\frac{1}{2})}}=\sqrt{\frac{n!}{\Gamma(n+\lambda+\frac{1}{2})}}(-1)^{n}|x|^\lambda e^{-\frac{x^2}{2}}L^{\lambda-\frac{1}{2}}_{n}(x^2),\\
&&\hspace{-27mm}\psi^{\lambda}_{2n+1}(x):=\frac{|x|^\lambda
e^{-\frac{x^2}{2}}H^{\lambda}_{2n+1}(x)}{\sqrt{2^{4n+2}
n!\Gamma(n+\lambda+\frac{3}{2})}}=\sqrt{\frac{n!}{\Gamma(n+\lambda+\frac{3}{2})}}(-1)^{n}x|x|^{\lambda}
e^{-\frac{x^2}{2}}L^{\lambda+\frac{1}{2}}_{n}(x^2),\end{eqnarray}
\textbf{we remark that}
\begin{eqnarray}&&\hspace{-24mm}\forall \lambda > -\frac{1}{2}, \forall n\geq 0, \psi^{\lambda}_{2n} \in L^{2}(\mathbb{R}, dx)\nonumber
\end{eqnarray} \textbf{and}
\begin{eqnarray}&&\hspace{-24mm}\forall \lambda > -\frac{3}{2}, \forall n\geq 0, \psi^{\lambda}_{2n+1} \in L^{2}(\mathbb{R}, dx)\nonumber.\end{eqnarray}
\textbf{Furthermore the following $L^{2}(\mathbb{R}, dx)$-inner product relation holds}
\renewcommand\theequation{\arabic{tempeq}\alph{equation}}
\setcounter{equation}{-1} \addtocounter{tempeq}{1}
\begin{eqnarray}
&&\hspace{-14mm}\forall \lambda > -\frac{1}{2}, \forall n, m\geq 0, \langle \psi^{\lambda}_{n}|\psi^{\lambda}_{m}\rangle=\delta_{nm}.\end{eqnarray}
\textbf{Remark; For odd integers n and m the previous relation can be extended to $\lambda > -\frac{3}{2}$.\\
Therefore the $\{\psi^{\lambda}_{n}\}_{n\in \mathbb{N}}$ define an orthonormal set of $L^{2}(\mathbb{R}, dx)$, generating the Hilbert space $\mathcal{H}_{\lambda}=span\{\psi^{\lambda}_{n}|n\in \mathbb{N}\}\subset L^{2}(\mathbb{R}, dx)$ for $\lambda>-\frac{1}{2}$.} Therefore, they can be used to represent WHA, that will be discussed later.
\section{WHA, its representation and relationship with PHO model}
The Heisenberg algebra of $H_{3}$ can be extended by a reflection operator $\hat{R}$, then the WHA is generated as a unital algebra with the generators $\{1, \mathfrak{a}, \mathfrak{a}^{\dag}, \hat{R}\}$, which satisfy the (anti-)commutation relations
\renewcommand\theequation{\arabic{tempeq}\alph{equation}}
\setcounter{equation}{0} \addtocounter{tempeq}{1}
\begin{eqnarray}
&&\hspace{-14mm}[\mathfrak{a}, \mathfrak{a}^{\dag}]=1+2\lambda \hat{R},\\
&&\hspace{-14mm}\{\hat{R}, \mathfrak{a}\}=\{\hat{R},
\mathfrak{a}^{\dag}\}=0.
\end{eqnarray}
Here $\lambda$ is a real constant called Wigner parameter and $\hat{R}$ is Hermitian and unitary operator also possessing the properties
\renewcommand\theequation{\arabic{tempeq}\alph{equation}}
\setcounter{equation}{-1} \addtocounter{tempeq}{1}
\begin{eqnarray}
&&\hspace{-14mm}\hat{R}^{2}=I,
\hat{R}^{\dag}=\hat{R}^{-1}=\hat{R}.\end{eqnarray}
It also commutates with number operator $N$ that includes the eigenvector $|n, \lambda\rangle$, such that
\renewcommand\theequation{\arabic{tempeq}\alph{equation}}
\setcounter{equation}{-1} \addtocounter{tempeq}{1}
\begin{eqnarray}
&&\hspace{-14mm}N|n, \lambda\rangle= n|n,
\lambda\rangle.\end{eqnarray}
Here $N$ satisfies the following commutation relations
\renewcommand\theequation{\arabic{tempeq}\alph{equation}}
\setcounter{equation}{-1} \addtocounter{tempeq}{1}
\begin{eqnarray}
&&\hspace{-14mm}[N, \mathfrak{a}]=-\mathfrak{a}, [N,
\mathfrak{a}^{\dag}]=\mathfrak{a}^{\dag},\end{eqnarray} and
results a modified relation between the operators $\mathfrak{a}, \mathfrak{a}^{\dag}$ and number operator $N$ is established by
\renewcommand\theequation{\arabic{tempeq}\alph{equation}}
\setcounter{equation}{-1} \addtocounter{tempeq}{1}
\begin{eqnarray}
&&\hspace{-14mm}\mathfrak{a}^{\dag}\mathfrak{a}=N+\lambda(1-\hat{R}).\end{eqnarray}
It is well known that the functions $\psi^{\lambda}_{n}(x)$ represent the WHA, as follows
\renewcommand\theequation{\arabic{tempeq}\alph{equation}}
\setcounter{equation}{0} \addtocounter{tempeq}{1}
\begin{eqnarray}&&\hspace{-14mm}\mathfrak{a}|2n, \lambda\rangle=\sqrt{2n}|2n-1, \lambda\rangle,\\
&&\hspace{-14mm}\mathfrak{a}|2n+1,
\lambda\rangle=\sqrt{2n+2\lambda+1}|2n,
\lambda\rangle,\\
&&\hspace{-14mm}\mathfrak{a}^{\dag}|2n,
\lambda\rangle=\sqrt{2n+2\lambda+1}|2n+1,
\lambda\rangle,\\
&&\hspace{-14mm}\mathfrak{a}^{\dag}|2n+1,
\lambda\rangle=\sqrt{2n+2}|2n+2, \lambda\rangle,\end{eqnarray} and
\renewcommand\theequation{\arabic{tempeq}\alph{equation}}
\setcounter{equation}{-1} \addtocounter{tempeq}{1}
\begin{eqnarray}
&&\hspace{-14mm} \hat{R}|n, \lambda\rangle= (-1)^n|n,
\lambda\rangle.
\end{eqnarray}
Since $\hat{R}$ is the reflection operator, then the explicit differential forms of the generators $\mathfrak{a}, \mathfrak{a}^{\dag}$ are obtained \cite{Yang} as follows:
\renewcommand\theequation{\arabic{tempeq}\alph{equation}}
\setcounter{equation}{0} \addtocounter{tempeq}{1}
\begin{eqnarray}&&\hspace{-14mm}\mathfrak{a}=\frac{1}{\sqrt{2}}\left(\frac{d}{dx}+x-\frac{\lambda}{x}\hat{R}\right),\\
&&\hspace{-14mm}\mathfrak{a}^{\dag}=\frac{1}{\sqrt{2}}\left(-\frac{d}{dx}+x+\frac{\lambda}{x}\hat{R}\right),\end{eqnarray}
which provide us with the coordinate representation for the position $\hat{x}$ and its $\lambda-$deformed canonical pair $\hat{p}_{\lambda}$ as bellow
\renewcommand\theequation{\arabic{tempeq}\alph{equation}}
\setcounter{equation}{0} \addtocounter{tempeq}{1}
\begin{eqnarray}&&\hspace{-14mm}\hat{x}=\frac{\mathfrak{a}+\mathfrak{a}^{\dag}}{\sqrt{2}},\\
&&\hspace{-14mm}\hat{p}_{\lambda}=\frac{\mathfrak{a}-\mathfrak{a}^{\dag}}{i\sqrt{2}}=-i\frac{d}{dx}+i\frac{\lambda}{x}\hat{R}.\end{eqnarray}
It should be noticed that the representation for the WHA is unitary since not only the operators $\mathfrak{a}$ and $\mathfrak{a}^{\dag}$ are Hermitian conjugate of each other \textbf{but also the operators $\hat{R}$ and $N$ are Hermitian with respect to the inner product of $L^{2}(\mathbb{R}, dx)$}.

Let us remember the interesting connection between the WHA and the PHO model, which is expressed in the symmetric form in terms of the mutually adjoint operators
$\mathfrak{a}$ and $\mathfrak{a}^{\dag}$
\renewcommand\theequation{\arabic{equation}}
\setcounter{equation}{\value {tempeq}}
\begin{eqnarray}
&&\hspace{-21mm}H_{\lambda}=\frac{1}{2}\{\mathfrak{a}, \mathfrak{a}^{\dag}\},\nonumber\\
&&\hspace{-14mm}=\frac{1}{2}\left[-\frac{d^2}{dx^2}+x^2+\frac{\lambda(\lambda-1)}{x^2}\right].\end{eqnarray}
Meanwhile, in Refs. \cite{Dodonov, dehghani1, sasaki}, it has been shown that the second-order differential operators
\setcounter{tempeq}{\value{equation}}
\renewcommand\theequation{\arabic{tempeq}\alph{equation}}
\setcounter{equation}{0} \addtocounter{tempeq}{1}
\begin{eqnarray}
&&\hspace{-1.5cm} {J}_{+}^{\lambda}:=\frac{(\mathfrak{a}^{\dag})^2}{2}=\frac{1}{4}\left[\left(x-\frac{d}{dx}\right)^2-\frac{\lambda(\lambda-1)}{x^2}\right],\\
&&\hspace{-1.5cm}{J}_{-}^{\lambda}:=\frac{\mathfrak{a}^2}{2}=\frac{1}{4}\left[\left(x+\frac{d}{dx}\right)^2-\frac{\lambda(\lambda-1)}{x^2}\right],\\
&&\hspace{-1.5cm}{J}_{3}^{\lambda}:=\frac{H^{\lambda}}{2}=\frac{1}{4}\left[-\frac{d^2}{dx^2}+x^2+\frac{\lambda(\lambda-1)}{x^2}\right],
\end{eqnarray}
constitute the $su(1,1)$ Lie algebra as follows
\renewcommand\theequation{\arabic{tempeq}\alph{equation}}
\setcounter{equation}{-1} \addtocounter{tempeq}{1}
\begin{eqnarray}
&&\hspace{-1.5cm}\left[{J}_{+}^{\lambda},{J}_{-}^{\lambda}\right]=-2{J}_{3}^{\lambda},
\hspace{20mm}\left[{J}_{3}^{\lambda},{J}_{\pm}^{\lambda}\right]=\pm{J}_{\pm}^{\lambda}.
\end{eqnarray}
It is useful to stress that two operators ${J}_{+}^{\lambda}$ and ${J}_{-}^{\lambda}$ are Hermitian conjugate of each other with respect to the inner product (6) and ${J}_{3}^{\lambda}$ as well as $H^{\lambda}$ are self-adjoint operators.
\section{Wigner Cat States}
Here, we are interested in extending the notion of even and odd coherent states to potentials other than the harmonic oscillator. Then, we introduce Wigner cat states( WCS) as generalized even and odd coherent states (deformation of usual Schr\"{o}dinger cat states) of quantized fields:
\renewcommand\theequation{\arabic{tempeq}\alph{equation}}
\setcounter{equation}{0} \addtocounter{tempeq}{1}
\begin{eqnarray}&&\hspace{-1.5cm}|W\rangle_{\lambda, +}:=\sqrt{\frac{\left(\frac{|w|}{\sqrt{2}}\right)^{2\lambda-1}}{I_{\lambda-\frac{1}{2}}(|w|^2)}}\sum^{\infty}_{n=0}{\frac{w^{2n}}{\sqrt{2^{2n}n!\Gamma(n+\lambda+\frac{1}{2})}}}|2n, \lambda\rangle,\\
&&\hspace{-1.5cm}|W\rangle_{\lambda,
-}:=\sqrt{\frac{\left(\frac{|w|}{\sqrt{2}}\right)^{2\lambda-1}}{I_{\lambda+\frac{1}{2}}(|z|^2)}}\sum^{\infty}_{n=0}{\frac{w^{2n+1}}{\sqrt{2^{2n+1}n!\Gamma(n+\lambda+\frac{3}{2})}}}|2n+1,
\lambda\rangle,
\end{eqnarray}
where $I_{\lambda}(x)$ refers to the modified Bessel function of the first type \cite{Grad}, with a convergency radius of infinity that has been used in order to normalize the WCSs to unity i.e. $_{\lambda, \pm}\langle W|W\rangle_{\lambda, \pm}=1$ for $w\in \mathbb{C}$. \textbf{It is worth mentioning that the states $|W\rangle_{\lambda, \pm}$ are defined for $\lambda>-\frac{1}{2}$ and for $\lambda>-\frac{3}{2}$, respectively, due to: (i) the $\Gamma(.)$ coefficients involved in the definitions of $|W\rangle_{\lambda, \pm}$, and (ii) to the $L^2$ behavior of the $\psi^{\lambda}_{n}$ studied in Section 2}\\
The algebras involved in the WCSs turn out to be WHA and it's generalization to the $su(1,1)$ Lie algebra. In other words, they can be viewed as generalized even and odd (or generalized two-photon) coherent states and satisfy the same eigenvalue equations with the same eigenvalues, i.e.
\renewcommand\theequation{\arabic{tempeq}\alph{equation}}
\setcounter{equation}{0} \addtocounter{tempeq}{1}
\begin{eqnarray}
&&\hspace{-1.5cm}\mathfrak{a}^2|W\rangle_{\lambda, +}=w^2|W\rangle_{\lambda, +},\\
&&\hspace{-1.5cm}\mathfrak{a}^2|W\rangle_{\lambda,
-}=w^2|W\rangle_{\lambda, -}.
\end{eqnarray}It is found that the WCSs reduce to the Schr\"{o}dinger's cat states in (1) in a certain limit, i.e. $\lambda= 0$. In addition, these are proportional to the Barut-Giradello coherent states for the reduced part of the two particle Calegero-Sutherland model, discussed in Refs. \cite{Deh1, Deh2}.
\subsection{Statistical properties of WCSs}
In this subsection, we will set up detailed studies on statistical properties of constructed WCSs. This will be achieved by investigation of some of the features including the Mandel's parameter and squeezing \textbf{factors}.\\\\
$\diamondsuit${\em{\textbf{Anti-bunching effect and sub-Poissonian statistics}}}\\
Now we are in a position to study the anti-bunching effect as well as the statistics of $|W\rangle_{\lambda, \pm}$ given by equations (19). For this reason we employ Mandel's parameters $Q_{\pm}$ to describe the photon number fluctuations\footnote{If $Q < 0 (> 0)$,
the field is called sub(super)-Poissonian. Also $Q= 0$ corresponds
to the canonical coherent state.}, which are defined as
\begin{figure}
\begin{center}
\epsfig{figure=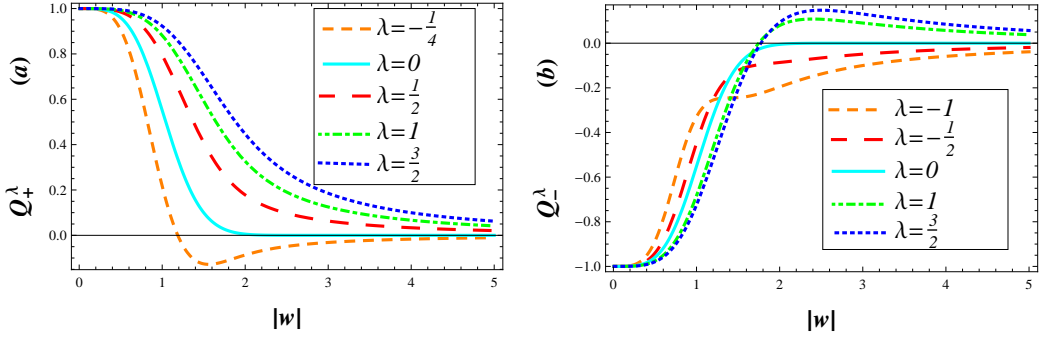,width=15cm}
\end{center}
\caption{\footnotesize The plots of  Mandel's parameters (a) ${Q}^{\lambda}_{+}$ and (b) ${Q}^{\lambda}_{-}$, as functions of $|{w}|$ for different values of $\lambda$.}
\end{figure}
\renewcommand\theequation{\arabic{equation}}
\setcounter{equation}{\value {tempeq}}
\begin{eqnarray}
&&\hspace{-14mm}Q_{\pm}^{\lambda}=
\frac{\langle{{N}}^2\rangle_{\pm}-{\langle{{N}\rangle}}_{\pm}^{2}
}{{\langle{{N}\rangle}}_{\pm}}-1.
\end{eqnarray}
For the generalized even and odd coherent state given by Eqs. (19) we have, respectively, \setcounter{tempeq}{\value{equation}}
\renewcommand\theequation{\arabic{tempeq}\alph{equation}}
\setcounter{equation}{0} \addtocounter{tempeq}{1}
\begin{eqnarray}
&&\hspace{-2.7cm}\langle{{N}}^2\rangle_{+}={\frac{(\frac{|w|}{\sqrt{2}})^{2\lambda-1}}{I_{\lambda-\frac{1}{2}}(|w|^2)}}\sum^{\infty}_{n=0}{\frac{|w|^{4n}4n^2}{{2^{2n}n!(n+\lambda-\frac{1}{2})!}}}=|w|^4-(2\lambda-1)|w|^{2}\frac{I_{\lambda+\frac{1}{2}}(|w|^2)}{I_{\lambda-\frac{1}{2}}(|w|^2)}\\
&&\hspace{-2.7cm}{\langle{{N}\rangle}}_{+}={\frac{(\frac{|w|}{\sqrt{2}})^{2\lambda-1}}{I_{\lambda-\frac{1}{2}}(|w|^2)}}\sum^{\infty}_{n=0}{\frac{|w|^{4n}2n}{{2^{2n}n!(n+\lambda-\frac{1}{2})!}}}=|w|^{2}\frac{I_{\lambda+\frac{1}{2}}(|w|^2)}{I_{\lambda-\frac{1}{2}}(|w|^2)}.
\end{eqnarray}
and
\renewcommand\theequation{\arabic{tempeq}\alph{equation}}
\setcounter{equation}{0} \addtocounter{tempeq}{1}
\begin{eqnarray}
&&\hspace{-2cm}\langle{{N}}^2\rangle_{-}={\frac{(\frac{|w|}{\sqrt{2}})^{2\lambda-1}}{I_{\lambda+\frac{1}{2}}(|w|^2)}}\sum^{\infty}_{n=0}{\frac{|w|^{4n+2}(2n+1)^2}{{2^{2n+1}n!(n+\lambda+\frac{1}{2})!}}}=|w|^4+4\lambda^{2}-(2\lambda-1)|w|^{2}\frac{I_{\lambda-\frac{1}{2}}(|w|^2)}{I_{\lambda+\frac{1}{2}}(|w|^2)}\\
&&\hspace{-2cm}{\langle{{N}\rangle}}_{-}={\frac{(\frac{|w|}{\sqrt{2}})^{2\lambda-1}}{I_{\lambda+\frac{1}{2}}(|w|^2)}}\sum^{\infty}_{n=0}{\frac{|w|^{4n+2}(2n+1)}{{2^{2n+1}n!(n+\lambda+\frac{1}{2})!}}}=|w|^{2}\frac{I_{\lambda-\frac{1}{2}}(|w|^2)}{I_{\lambda+\frac{1}{2}}(|w|^2)}-2\lambda.
\end{eqnarray}
From figure 1, the anti-bunching of the generalized even and odd coherent states is changed by variations of the Wigner parameter $\lambda$ as well as $|w|$. It is understood that the even WCSs, $|W\rangle_{\lambda, +}$ illustrate fully super-Poissonian for any $\lambda\geq0$, except for negative values of $-\frac{1}{2}<\lambda<0$ and $|w|>1$ where sub-Poissonian statistics emerges (see figure 1(a)). Figure 1(b) shows that the odd WCSs, $|W\rangle_{\lambda, -}$, are fully sub-Poissonian for any $-\frac{3}{2}<\lambda\leq0$. Although, $|W\rangle_{\lambda, -}$ become super-Poissonian for $\lambda>0$ when $|w|$ increases.\\\\
$\diamondsuit${\em{\textbf{Squeezing effect in field operators $\hat{x}$ and $\hat{p}_{\lambda}$}}}\\
From the commutation relation
\renewcommand\theequation{\arabic{equation}}
\setcounter{equation}{\value {tempeq}}
\begin{eqnarray}
&&\hspace{-14mm}[\hat{x}, \hat{p}_{\lambda}]=i(1+2\lambda \hat{R})
\end{eqnarray}
the uncertainty relation for the variances of the operators $\hat{x}$ and $\hat{p}_{\lambda}$\footnote{Without \textbf{loss} of generality, we have set $\hat{p}_{\lambda}$ by $\hat{p}$ for the remainder of this work.} can be calculated to be
\begin{eqnarray}
&&\hspace{-14mm}\langle\sigma_{xx}\rangle \langle\sigma_{pp}\rangle
\geq \frac{|\langle1+2\lambda \hat{R}\rangle|^{2}}{4},
\end{eqnarray}
where $\langle\sigma_{\hat{x}\hat{y}}\rangle= \frac{\langle \hat{x}\hat{y}+{\hat{y}}\hat{x}\rangle}{2}-\langle \hat{x}\rangle\langle \hat{y}\rangle$ and the angular brackets denote averaging over an arbitrary normalizable state for which the mean values are well defined, $\langle \hat{y}\rangle={_{\pm, \lambda}\langle W|}\hat{y}|W\rangle_{\lambda, \pm}$. Following Walls (1983) as well as Wodkiewicz (1985) \cite{WODKIEWICZ, Walls} it can be said that a state is squeezed if the condition $\langle\sigma_{xx}\rangle < \frac{|\langle1+2\lambda \hat{R}\rangle|}{2}$ or $\langle\sigma_{pp}\rangle < \frac{|\langle1+2\lambda \hat{R}\rangle|}{2}$ is fulfilled. In other words, a set of quantum states are called squeezed states if they have less uncertainty in one parameter ($\hat{x}$ or $\hat{p}$) than coherent states. Then to measure the degree of squeezing we introduce the squeezing factors $S_{x(p),+}$ and $S_{x(p),-}$ \cite{Buzek1}, corresponding with the even and odd WCSs, respectively
\setcounter{tempeq}{\value{equation}}
\renewcommand\theequation{\arabic{tempeq}\alph{equation}}
\setcounter{equation}{0} \addtocounter{tempeq}{1}
\begin{eqnarray}
&&\hspace{-14mm}S_{x,\pm}=\frac{_{\pm}\langle\sigma_{xx}\rangle_{\pm}-\frac{|_{\pm}\langle1+2\lambda
\hat{R}\rangle_{\pm}|}{2}}{\frac{|_{\pm}\langle1+2\lambda
\hat{R}\rangle_{\pm}|}{2}},\\
&&\hspace{-14mm}S_{p,\pm}=\frac{_{\pm}\langle\sigma_{pp}\rangle_{\pm}-\frac{|_{\pm}\langle1+2\lambda
\hat{R}\rangle_{\pm}|}{2}}{\frac{|_{\pm}\langle1+2\lambda
\hat{R}\rangle_{\pm}|}{2}},
\end{eqnarray}
which results that the squeezing condition takes the simple form of $S_{x(p),\pm}< 0$. By using the mean values of the generators of the WHA, one can derive the variance and covariance of the operators $\hat{x}$ and $\hat{p}$ corresponding with the states $|W\rangle_{\lambda, \pm}$ as follow
\renewcommand\theequation{\arabic{tempeq}\alph{equation}}
\setcounter{equation}{0} \addtocounter{tempeq}{1}
\begin{eqnarray}
&&\hspace{-2.7cm}\langle\sigma_{xx}\rangle_{+}:=\langle\hat{x}^2\rangle_{+}-\langle\hat{x}\rangle^2_{+}=2(|w|\cos{\phi})^2-|w|^2+\lambda+\frac{1}{2}+|w|^2\frac{I_{\lambda+\frac{1}{2}}(|w|^2)}{I_{\lambda-\frac{1}{2}}(|w|^2)}\\
&&\hspace{-2.7cm}\langle\sigma_{pp}\rangle_{+}:=\langle\hat{p}^2\rangle_{+}-\langle\hat{p}\rangle^2_{+}=-2(|w|\cos{\phi})^2+|w|^2+\lambda+\frac{1}{2}+|w|^2\frac{I_{\lambda+\frac{1}{2}}(|w|^2)}{I_{\lambda-\frac{1}{2}}(|w|^2)}\\
&&\hspace{-2.7cm}\langle\sigma_{xx}\rangle_{-}:=\langle\hat{x}^2\rangle_{-}-\langle\hat{x}\rangle^2_{-}=2(|w|\cos{\phi})^2-|w|^2-\lambda+\frac{1}{2}+|w|^2\frac{I_{\lambda-\frac{1}{2}}(|w|^2)}{I_{\lambda+\frac{1}{2}}(|w|^2)}\\
&&\hspace{-2.7cm}\langle\sigma_{pp}\rangle_{-}:=\langle\hat{p}^2\rangle_{-}-\langle\hat{p}\rangle^2_{-}=-2(|w|\cos{\phi})^2+|w|^2-\lambda+\frac{1}{2}+|w|^2\frac{I_{\lambda-\frac{1}{2}}(|w|^2)}{I_{\lambda+\frac{1}{2}}(|w|^2)}
\end{eqnarray}
\begin{figure}
\begin{center}
\epsfig{figure=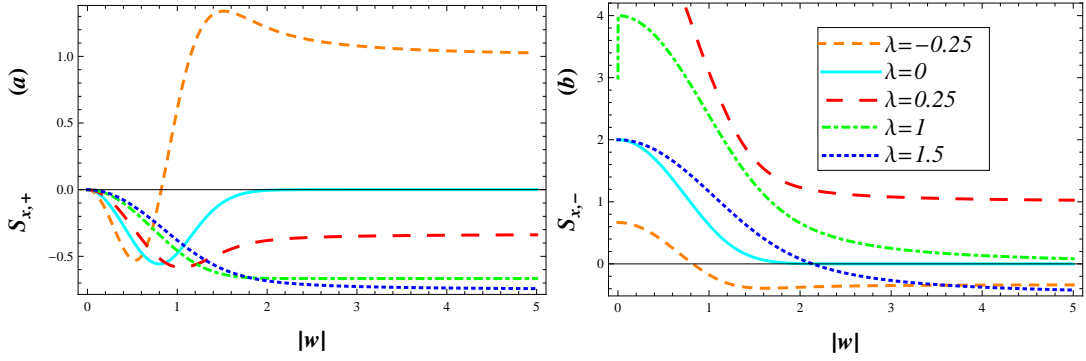,width=15cm}
\end{center}
\caption{\footnotesize The plots of squeezing factors (a) $S_{x,+}$ and (b) $S_{x,-}$ in terms of $|w|$ for different values of $\lambda$ while we choose the phase $\phi=\frac{\pi}{2}$.}
\end{figure}

\begin{figure}
\begin{center}
\epsfig{figure=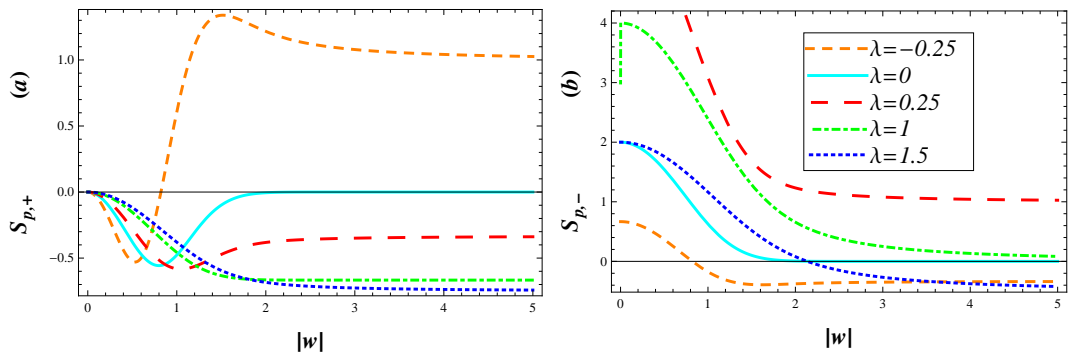,width=15cm}
\end{center}
\caption{\footnotesize The plots of squeezing factors (a) $S_{p,+}$ and (b) $S_{p,-}$ in terms of $|w|$ for different values of $\lambda$ while we choose the phase $\phi=0$.}
\end{figure}

\begin{figure}
\begin{center}
\epsfig{figure=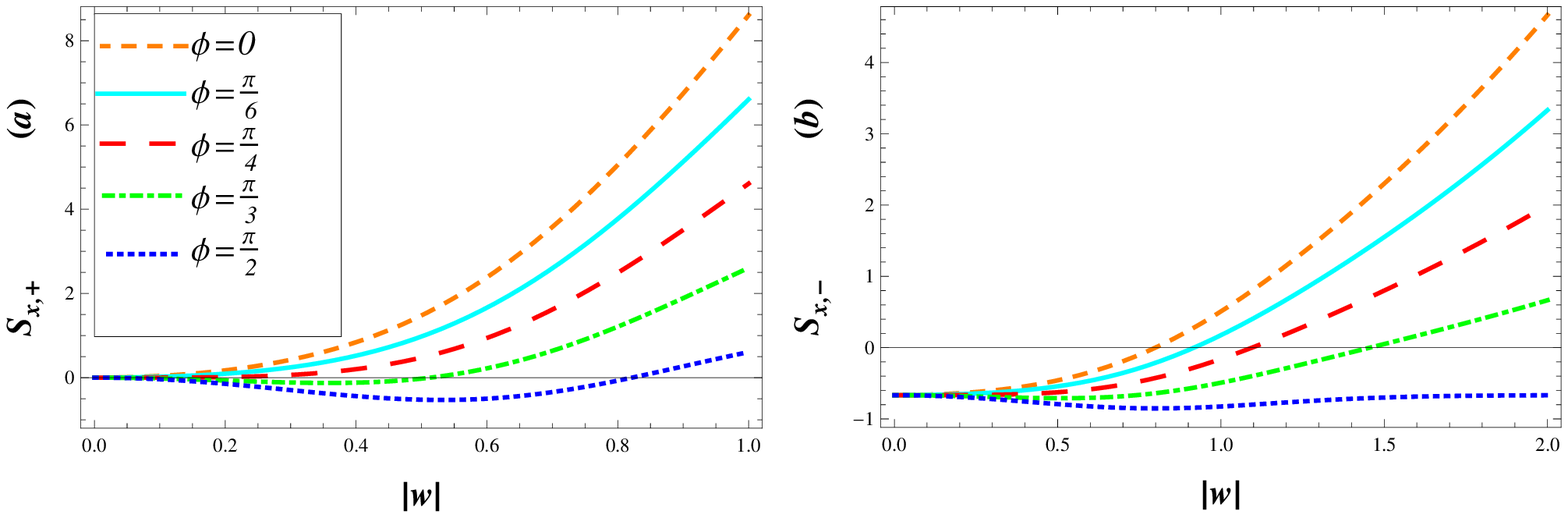,width=15cm}
\end{center}
\caption{\footnotesize The plots of squeezing factors (a) $S_{x,+}$ and (b) $S_{x,-}$ in terms of $|w|$ for different values of the phase $\phi$ while we choose $\lambda=-\frac{1}{4}$ and $\lambda=-1$, respectively.}
\end{figure}
\textbf{It results that $S_{x(p),\pm}$ is strongly dependent on the complex variable $w(= |w| e^{i\phi})$ and the deformation parameter $\lambda$. These dependencies can be discussed as follows:}\\
$\bullet$ For the case $\phi=\frac{\pi}{2}$, our calculations show that the squeezing factors $S_{x,\pm}$ are really dependent of $\lambda$. As shown in figure 2(a), squeezing factor $S_{x,+}$ becomes negative for any values of $|w|$ while $\lambda>0$, which indicates that the states $|W\rangle_{\lambda,+}$ exhibit squeezing effects in the position coordinate $\hat{x}$. We find that by increasing $\lambda$, the degree of squeezing is enhanced. However, for negative values of $\lambda$, $-\frac{1}{2}<\lambda<0$, $S_{x,+}$ becomes negative in small values of $|w|$ and we will loss the squeezing when $|w|$ is increasing. Figure 2(b) shows that, in contrast with the even WCSs, the odd WCSs become squeezed for negative values of $-\frac{3}{2}<\lambda<0$ and by decreasing $\lambda$ the degree of squeezing or depth of non-classicality is enhanced. However we will loss the squeezing in $\hat{x}$ when $\lambda\geq0$.\\\\
$\bullet$ In figure 3, we have plotted $S_{p,\pm}$ versus $|w|$ for different values of $\lambda$ when $\phi$ tends to zero. \textbf{It implies that squeezing in the $p$ component passes the greatest value when $\phi$ reaches zero, where squeezing in the position coordinate, $x$, is disappeared.}\\\\
$\bullet$ Figure 4 visualizes variation of the squeezing factors $S_{x,+}$ and $S_{x,-}$ for different values of the phase $\phi$ when we choose the deformed parameter $\lambda=-0.25$ and $\lambda=-1$, respectively. \textbf{It is worth mentioning that the squeezing effect in the field operator $\hat{x}$ are
considerable for the case $\phi\geq\frac{\pi}{3}$. Figure 4(b) shows that the squeezing factor $S_{x,-}$ becomes smaller than zero for $\lambda=-1$ and for any values of $\phi$}.\\
$\diamondsuit${\em{\textbf{Higher order squeezing}}}\\ We introduce two generalized Hermitian quadrature operators $X_{1}$ and $X_{2}$ as
\renewcommand\theequation{\arabic{tempeq}\alph{equation}}
\setcounter{equation}{-1} \addtocounter{tempeq}{1}
\begin{eqnarray}
&&\hspace{-2.7cm}X_{1}:=\frac{J^{\lambda}_{-}+J^{\lambda}_{+}}{2},\hspace{3mm}X_{2}:=\frac{J^{\lambda}_{-}-J^{\lambda}_{+}}{2i}
\end{eqnarray}
with the commutation relation $[X_{1} ,X_{2} ] = iJ^{\lambda}_{3}$.
From this commutation relation the uncertainty relation for the
variances of the quadrature operators $X_{i}$ follows
\renewcommand\theequation{\arabic{tempeq}\alph{equation}}
\setcounter{equation}{-1} \addtocounter{tempeq}{1}
\begin{eqnarray}
&&\hspace{-2.7cm}\langle(\Delta X_{1})^2\rangle\langle(\Delta X_{2})^2\rangle\geq\frac{|\langle J^{\lambda}_{3}\rangle|^2}{4} ,\end{eqnarray} where $\langle(\Delta X_{i})^2\rangle=\langle X_{i})^2\rangle-\langle X_{i}\rangle^2$ and the angular brackets denote averaging over the even or odd WCSs. To analyze the higher order or $\rm su(1, 1)$ squeezing, we introduce the squeezing factors\footnote{\textbf{Squeezing is said to exist if} $-1\leq S_{1(2),\pm}<0$.} $S_{1(2),+}$ and $S_{1(2),-}$ corresponding with even and odd WCSs, respectively, as
\renewcommand\theequation{\arabic{tempeq}\alph{equation}}
\setcounter{equation}{0} \addtocounter{tempeq}{1}
\begin{eqnarray}
&&\hspace{-14mm}S_{1,\pm}:=\frac{_{\pm}\langle(\Delta
X_{1})^2\rangle_{\pm}-\frac{|_{\pm}\langle
J^{\lambda}_{3}\rangle_{\pm}|}{2}}{\frac{|_{\pm}\langle J^{\lambda}_{3}\rangle_{\pm}|}{2}},\\
&&\hspace{-14mm}S_{2,\pm}:=\frac{_{\pm}\langle(\Delta
X_{2})^2\rangle_{\pm}-\frac{|_{\pm}\langle
J^{\lambda}_{3}\rangle_{\pm}|}{2}}{\frac{|_{\pm}\langle
J^{\lambda}_{3}\rangle_{\pm}|}{2}}.\end{eqnarray} For calculation of the squeezing degree in the fields $X_{1(2)}$ one can use mean values of generators of the $su(1, 1)$ Lie algebra as
\renewcommand\theequation{\arabic{tempeq}\alph{equation}}
\setcounter{equation}{0} \addtocounter{tempeq}{1}
\begin{eqnarray}
&&\hspace{-2.7cm}_{+}\langle(\Delta
X_{1})^2\rangle_{+}:=\langle{X_{1}}^2\rangle_{+}-\langle X_{1}\rangle^2_{+}=\frac{\lambda}{4}+\frac{1}{8}+|w|^2\frac{I_{\lambda+\frac{1}{2}}(|w|^2)}{4I_{\lambda-\frac{1}{2}}(|w|^2)}\\
&&\hspace{-2.7cm}_{+}\langle(\Delta
X_{2})^2\rangle_{+}:=\langle{X_{2}}^2\rangle_{+}-\langle X_{2}\rangle^2_{+}=\frac{\lambda}{4}+\frac{1}{8}+|w|^2\frac{I_{\lambda+\frac{1}{2}}(|w|^2)}{4I_{\lambda-\frac{1}{2}}(|w|^2)}\\
&&\hspace{-2.7cm}_{+}\langle J^{\lambda}_{3}\rangle_{+}=\frac{\lambda}{2}+\frac{1}{4}+|w|^2\frac{I_{\lambda+\frac{1}{2}}(|w|^2)}{2I_{\lambda-\frac{1}{2}}(|z|^2)}\\
&&\hspace{-2.7cm}_{-}\langle(\Delta
X_{1})^2\rangle_{-}:=\langle{X_{1}}^2\rangle_{-}-\langle X_{1}\rangle^2_{-}=-\frac{\lambda}{4}+\frac{1}{8}+|w|^2\frac{I_{\lambda-\frac{1}{2}}(|w|^2)}{4I_{\lambda+\frac{1}{2}}(|w|^2)}\\
&&\hspace{-2.7cm}_{-}\langle(\Delta
X_{2})^2\rangle_{-}:=\langle{X_{2}}^2\rangle_{-}-\langle
X_{2}\rangle^2_{-}=-\frac{\lambda}{4}+\frac{1}{8}+|w|^2\frac{I_{\lambda-\frac{1}{2}}(|w|^2)}{4I_{\lambda+\frac{1}{2}}(|w|^2)}\\
&&\hspace{-2.7cm}_{+}\langle
J^{\lambda}_{3}\rangle_{-}=-\frac{\lambda}{2}+\frac{1}{4}+|w|^2\frac{I_{\lambda-\frac{1}{2}}(|w|^2)}{2I_{\lambda+\frac{1}{2}}(|w|^2)}.\end{eqnarray}
Clearly, by means of numerical calculations, from Eqs. (31), we conclude that even and odd WCSs do not exhibit the possibility of squeezing effect neither in $X_{1}$ nor in $X_{2}$ for any values of $\lambda, w$, i.e.
\renewcommand\theequation{\arabic{tempeq}\alph{equation}}
\setcounter{equation}{-1} \addtocounter{tempeq}{1}
\begin{eqnarray}
&&\hspace{-14mm}S_{1(2),\pm}=0.\end{eqnarray}
Therefore, we ask whether even and odd WCSs can be good candidates for the states which minimize the uncertainty conditions (29) or not?. \textbf{Our calculations show that the minimum uncertainty condition is hold, i.e.}
\renewcommand\theequation{\arabic{tempeq}\alph{equation}}
\setcounter{equation}{-1} \addtocounter{tempeq}{1}
\begin{eqnarray}
&&\hspace{-2.7cm}\langle(\Delta X_{1})^2\rangle\langle(\Delta
X_{2})^2\rangle=\frac{|\langle
J^{\lambda}_{3}\rangle|^2}{4}.\end{eqnarray}
\section{Conclusions}
To sum up, we introduced a new family of generalized even and odd coherent states (or Wigner cat states), which include rather different statistical properties than the well known Schr\"{o}dinger cat states. In contrast with the Schr\"{o}dinger cat states\footnote{Detailed studies on the statistical properties of the Schr\"{o}dinger's cat states show that the even CSs obey Poissonian statistics and have squeezing effects. However, the odd CSs exhibit sub-Poisonian statistics and do not include squeezing effects.}, the anti-bunching effect as well as the squeezing properties appear for both even and odd WCSs. For instance, fully anti-bunching effect ( or fully sub-poissonian statistics) only can be observed in the odd WCSs. Although, the anti-bunching effect occurs in the even WCSs for negative values of the parameter $\lambda$ as well as large values of $|w|$. For given values of the phase $\phi$($=\frac{\pi}{2}$) as well as positive values of the Wigner parameter $\lambda$, fully squeezing features occur in the even WCSs. Also, fully squeezing behavior occur in the odd WCSs for negative values of the Wigner parameter $\lambda$ when we choose the phase $\phi=\frac{\pi}{2}$. The amplitude-squared squeezing and anti-bunching effect disappear for both even and odd WCSs for all values\footnote{All values of the parameters $\lambda, \phi$ and $|w|$, that were chosen here, are optional.} of the parameters $\lambda, \phi$ and $|w|$. It should be mentioned that the states $|W\rangle_{\pm}$ minimize the uncertainty relations of each pair of the $su(1,1)$ components $X_{1}$ and $X_{2}$. It is interesting to note that when $\lambda\rightarrow0$, the even and odd WCSs become the usual Schrodinger cat states. Therefore, the usual cat states are the special cases of the WCSs.\\\\
\textbf{Acknowledgements}\\
The first author would like to thank the reviewer for thorough reviews and, in particular, for constructive critique and expert suggestions for improvement the presentation.

\end{document}